\documentclass[12pt]{article}
\usepackage{graphicx}
\newcommand{\interlinia}{}

\title{Statistical theory of elastic constants
of cholesteric liquid crystals}

\author{A. Kapanowski \\
{\em Institute of Physics, Jagellonian University,}\\
{\em ulica Reymonta 4, 30-059 Cracow, Poland}  }

\begin{document}
\maketitle

\begin{abstract}
\interlinia
A statistical theory of cholesteric liquid crystals composed 
of short rigid biaxial molecules is presented. 
It is derived in the thermodynamic
limit at a small density and a small twist. 
The uniaxial (biaxial) cholesteric phase is regarded as a distorted 
form of the uniaxial (biaxial) nematic phase.
The chirality of the interactions and the implementation of the 
inversion to the rotation matrix elements are discussed in detail.
General microscopic expressions for the elastic constants are derived.
The expressions involve the one-particle distribution function and
the potential energy of two-body short-range interactions.
It is shown that the elastic constants determine the twist 
of the phase.
The stability condition for the cholesteric and nematic
phases is presented.

The theory is used to study unary and binary systems.
The temperature and concentration dependence of the
order parameters, the elastic constants and the twist 
of the phase are obtained.
The possibility of phase separation is not investigated.

\end{abstract}

\interlinia

\section{Introduction}

The cholesteric phase can be considered as a special case 
of the nematic phase 
\cite{[1993_de_Gennes]}. 
The long axes of the anisotropic molecules
are on the average aligned parallel to each other within planes.
The direction of this alignment rotates smoothly as one proceeds
in a direction perpendicular to the parallel planes.
Such a phase (the twisted uniaxial nematic phase)
will be called the uniaxial cholesteric phase.
By analogy, the twisted biaxial nematic phase will be called
the biaxial cholesteric phase. We note that in the case of the twisted
biaxial nematic phase three twists around three perpendicular axes
of biaxial symmetry should be considered simultaneously.

There are many models of the uniaxial cholesteric phase where 
molecules are assumed to be uniaxial. 
In 1970 Goossens 
\cite{[1970_Goossens]} 
showed within the extended Maier-Saupe model 
that the quadrupole interactions give rise to the twist.
In 1976 Straley 
\cite{[1976_Straley]} 
presented a statistical-mechanical theory of the elastic constants 
and of the spontaneous twisting of a cholesteric. 
He expressed the pitch by means of the elastic constants.
We note that he gave the geometric interpretation of the chiral term
which is often present in microscopic theories of cholesterics.
This term can approximately describe the minimum approach distance of
a pair of threaded rods. We will derive similar terms in the case of
biaxial molecules.
In 1977 Lin-Liu {\em et al.} 
\cite{[1977_Lin-Liu_Shih_Woo]}
presented a molecular theory of cholesteric liquid crystals and showed
the general form of the chiral potential energy of interactions for
uniaxial molecules. They discussed different types of the temperature
dependence of the pitch.

Real molecules forming liquid crystalline phases are never uniaxial and
it is more realistic to assume less symmetric biaxial molecules.
Some properties of the cholesteric phase formed by biaxial molecules
were investigated in the past.
In 1974 Priest and Lubensky 
\cite{[1974_Priest_Lubensky]} 
found the order of the biaxial order parameter and predicted 
a fluctuation instability. Van der Meer and Vertogen 
\cite{[1976_Meer_Vertogen]}
discussed the dependence of the pitch on the biaxial order parameter.
They also derived expressions for the temperature dependence 
of the pitch and the elastic constants 
\cite{[1979_Meer_Vertogen]}
in the case of the uniaxial cholesteric phases. Finally, in 1992 Evans 
\cite{[1992_Evans]}
presented a hard body model for chiral nematic liquid crystals.
The density functional theory was used, and molecules were represented
by a hard convex twisted ellipsoidal core, with and without 
an encircling isotropic square well. The pitch was found to be
density and temperature independent with values in the visible
region of the spectrum. Long range potential softness could account
for the increase of the pitch with decreasing temperature.

The phase behaviour of liquid crystalline mixtures has been studied, 
using a number of theoretical methods. In 1980 Sivardiere 
\cite{[1980_Sivaldiere]}
introduced the Ising-like model and obtained a large variety of phase
diagrams. Brochard {\em et al.} 
\cite{[1984_Brochard_Jouffroy_Levinson]}
considered the Maier-Saupe model and gave a cataloque of allowed diagrams
for mixtures of nematogens.
As far as the cholesteric mixtures are concerned, the theoretical and 
experimental works describe usually uniaxial cholesteric phases
composed of uniaxial molecules.
In 1971 Nakagiri {\em et al.} 
\cite{[1971_Nakagiri_Kodama_Kobayashi]}
studied the helical twisting power in 
nematic-cholesteric mixtures as a function of concentration. 
It was found that, as the concentration of the cholesteric material 
increases, the twisting power of the mixtures increases lineary 
up to a certain concentration, beyound which it increases more slowly, 
taking a maximum value at some specific concentration. 
Then, in the high-concentration region, the twisting power falls.
We note that typically the additivity rule is satisfied as described in
Sec. \ref{sec:corner}. In 1977 Lin-Liu {\em et al.} 
\cite{[1977_Lin-Liu_Shih_Woo]} 
presented a molecular theory of binary cholesteric mixtures. 
They considered uniaxial molecules and derived a formula showing how
the pitch depends on temperature and concentration. Under certain 
conditions the formula reduces at fixed
$T-T_{C}$ to a simple quadratic rational fraction in the composition.

The forming of the cholesteric phase is closely connected with
chirality. 
Some authors tried to establish quantitative relationships between
molecular properties and measurable properties that result from chiral
molecular structures. Osipov {\em et al.} 
\cite{[1995_Osipov_Pickup_Dunmur]}
proposed an intrinsic molecular chirality tensor based only on nuclear
position. The chirality tensor gives rise to two universal chirality
indices, the first givig information about absolute chirality,
and the second about anisotropy of the chirality.
Recently Harris {\em et al.} 
\cite{[1999_Harris_Kamien_Lubensky]}
showed that any chiral measure of a geometric object is a pseudoscalar
and must involve three-point correlations that only come into play
when the molecule has at least four atoms. In general, a molecule is 
characterized by an infinite set of chiral parameters.
However, one can also consider chirality of interactions
\cite{[1976_Lin-Liu_Shih_Woo_Tan],[1999_Issaenko_Harris_Lubensky]} 
and we will use this approach.

Our aim is to describe the uniaxial and biaxial cholesteric phases 
that consist of biaxial or uniaxial molecules. 
We would like to derive the 
microscopic expressions for the elastic constants and the pitch. 
We will investigate the inversion in the context of chirality. 
Our paper is organized as follows.
In Sec. \ref{sec:pheno} we present a phenomenological continuum theory 
of cholesteric liquid crystals.
In Sec. \ref{sec:micro} we describe a statistical theory of cholesteric 
phases that consist of rigid biaxial molecules, and 
in Sec. \ref{sec:elastic} we derive general expressions for the elastic 
constants and the cholesteric pitch. A condition of stability is 
obtained that concerns both nematics and cholesterics.
Exemplary calculations are presented in Sec. \ref{sec:corner}, 
were the Corner potential energy is applied. 
By means of symmetry considerations we will identyfy
main chiral terms for biaxial molecules. In the limit of uniaxial 
molecules we will recover the results from
\cite{[1977_Lin-Liu_Shih_Woo]}.
In Sec. \ref{sec:summary} we summarize the results of this work.

\section{Phenomenological approach}
\label{sec:pheno}

In this section we will describe a uniform phase from
a phenomenological point of view 
\cite{[1994_Stallinga_Vertogen]}. 
We assume that at every point $\vec{r}$ inside a considered phase 
we can define three orthonormal vectors
$(\vec{L}(\vec{r}),\vec{M}(\vec{r}),\vec{N}(\vec{r}))$
reflecting orientational properties of this phase. 
In the case of the biaxial phase
they determine directions of its two-fold axes of symmetry.
The vectors $(\vec{L},\vec{M},\vec{N})$ create the local frame 
which can be expressed by means of a space-fixed reference frame 
$(\vec{e}_{x},\vec{e}_{y},\vec{e}_{z})$ as
\begin{equation}
\vec{L} = L_{\alpha}\vec{e}_{\alpha},
\ \vec{M} = M_{\alpha}\vec{e}_{\alpha},
\ \vec{N} = N_{\alpha}\vec{e}_{\alpha},
\label{eq:LMN=Re}
\end{equation}
where repeated indices imply summation.
The completely ordered uniform phase is described by
$(\vec{L},\vec{M},\vec{N})=(\vec{e}_{x},\vec{e}_{y},\vec{e}_{z})$.

Let us call $F_{d}$ the free energy due to the distortion 
of the local frame $(\vec{L},\vec{M},\vec{N})$.
A general form of its density $f_{d}(\vec{r})$ was derived in
\cite{[1994_Stallinga_Vertogen]} 
in the case of small distortions.
When a considered phase has a $D_{2}$ symmetry group
(the biaxial cholesteric phase) we get
\begin{eqnarray}
\label{eq:d2}
f_{d} & = &
K_{11} D_{11} + K_{22} D_{22} + K_{33} D_{33}
\nonumber \\
& & +{\frac{1}{2}} K_{1111} (D_{11})^{2}
+{\frac{1}{2}} K_{1212} (D_{12})^{2}
+{\frac{1}{2}} K_{1313} (D_{13})^{2} 
\nonumber \\
& & +{\frac{1}{2}} K_{2121} (D_{21})^{2}
+{\frac{1}{2}} K_{2222} (D_{22})^{2}
+{\frac{1}{2}} K_{2323} (D_{23})^{2} 
\nonumber \\
& & +{\frac{1}{2}} K_{3131} (D_{31})^{2}
+{\frac{1}{2}} K_{3232} (D_{32})^{2}
+{\frac{1}{2}} K_{3333} (D_{33})^{2} 
\nonumber \\
& & + {\frac{1}{2}} (K_{1221}+K_{1122}) (D_{12}D_{21}+D_{11}D_{22})
\nonumber \\
& & + {\frac{1}{2}} (K_{1331}+K_{1133}) (D_{13}D_{31}+D_{11}D_{33})
\nonumber \\
& & + {\frac{1}{2}} (K_{2332}+K_{2233}) (D_{23}D_{32}+D_{22}D_{33})
\nonumber \\
& & + {\frac{1}{2}} (K_{1221}-K_{1122}) (D_{12}D_{21}-D_{11}D_{22})
\nonumber \\
& & + {\frac{1}{2}} (K_{1331}-K_{1133}) (D_{13}D_{31}-D_{11}D_{33})
\nonumber \\
& & + {\frac{1}{2}} (K_{2332}-K_{2233}) (D_{23}D_{32}-D_{22}D_{33})
\nonumber \\
& & + L_{123}\partial_{\alpha}(L_{\alpha}D_{23}+M_{\alpha}D_{13})
\nonumber \\
& & + L_{231}\partial_{\alpha}(M_{\alpha}D_{31}+N_{\alpha}D_{21})
\nonumber \\
& & + L_{312}\partial_{\alpha}(N_{\alpha}D_{12}+L_{\alpha}D_{32}),
\end{eqnarray}
where $K_{ij}$, $K_{ijkl}=K_{klij}$, $L_{ijk}=L_{jik}$ 
are the elastic constants,
\begin{eqnarray}
D_{11}=L_{\alpha} M_{\beta} \partial_{\alpha} N_{\beta}, &
D_{12}=L_{\alpha} N_{\beta} \partial_{\alpha} L_{\beta}, &
D_{13}=L_{\alpha} L_{\beta} \partial_{\alpha} M_{\beta},\nonumber \\
D_{21}=M_{\alpha} M_{\beta} \partial_{\alpha} N_{\beta}, &
D_{22}=M_{\alpha} N_{\beta} \partial_{\alpha} L_{\beta}, &
D_{23}=M_{\alpha} L_{\beta} \partial_{\alpha} M_{\beta},\nonumber \\
D_{31}=N_{\alpha} M_{\beta} \partial_{\alpha} N_{\beta}, &
D_{32}=N_{\alpha} N_{\beta} \partial_{\alpha} L_{\beta}, &
D_{33}=N_{\alpha} L_{\beta} \partial_{\alpha} M_{\beta}.
\end{eqnarray}
The terms with $K_{ii}$ give 3 bulk terms,
the terms with $L_{ijk}$ give 3 surface terms,
the terms with $K_{ijkl}$ give 12 bulk and 3 surface terms
of the form
\begin{eqnarray}
\partial_{\alpha}(L_{\beta}\partial_{\beta}L_{\alpha}-
L_{\alpha}\partial_{\beta}L_{\beta})
&=& 2 (D_{23}D_{32}-D_{22}D_{33}),
\nonumber\\
\partial_{\alpha}(M_{\beta}\partial_{\beta}M_{\alpha}-
M_{\alpha}\partial_{\beta}M_{\beta})
&=& 2 (D_{31}D_{13}-D_{11}D_{33}),
\nonumber\\
\partial_{\alpha}(N_{\beta}\partial_{\beta}N_{\alpha}-
N_{\alpha}\partial_{\beta}N_{\beta})
&=& 2 (D_{12}D_{21}-D_{11}D_{22}).
\end{eqnarray}
The total numbers of bulk and surface terms are 15 and 6, 
respectively.

If a considered phase possesses a $D_{\infty}$ symmetry group
(the uniaxial cholesteric phase),
the number of the elastic constants is smaller, because some
constants from the previous case become dependent or zero.
Let the $z$ axis be oriented along the axis of symmetry.
Then the distortion free-energy density has the form 
of the Frank expression
\begin{eqnarray}
\label{eq:d8}
f_{d} & = & K_{0} \vec{N} \cdot (\nabla \times \vec{N} )
+ {\frac {1}{2}} K_{1} (\nabla \cdot \vec{N} )^{2}
\nonumber \\
& + & {\frac {1}{2}} K_{2} [ \vec{N} \cdot (\nabla \times \vec{N} ) ]^{2}
+ {\frac {1}{2}} K_{3} [ \vec{N} \times (\nabla \times \vec{N} ) ]^{2} 
\nonumber \\
& + & {\frac {1}{2}} K_{4}\nabla \cdot [(\vec{N} \cdot \nabla ) \vec{N}
                         - \vec{N} (\nabla \cdot \vec{N})]
\nonumber \\
& + & {\frac {1}{2}} K_{5}\nabla \cdot [(\vec{N} \cdot \nabla ) \vec{N}
                + \vec{N} (\nabla \cdot \vec{N})],
\end{eqnarray}
where the relations among the nonzero elastic constants are
\begin{eqnarray}
& K_{0} = K_{11} = K_{22}, & \nonumber\\
& K_{1} = K_{1212} = K_{2121}, & \nonumber\\
& K_{2} = K_{1111} = K_{2222}, & \nonumber\\
& K_{3} = K_{3131} = K_{3232}, & \nonumber\\
& K_{4} = (K_{1221}-K_{1122}+K_{1}+K_{2})/2, & \nonumber\\
& K_{5} = 2L_{231} = -2L_{312}. &
\end{eqnarray}
Therefore in the case of the uniaxial phase we have 4 bulk
(from $K_{0}$ to $K_{3}$) and 2 surface terms ($K_{4}$ and $K_{5}$).

In the continuum theory of uniaxial nematic liquid crystals
three basic types of deformations i.e. splay, twist and bend, 
appear, which extract from the distortion 
free energy terms with $K_{1}$, $K_{2}$ and $K_{3}$, respectively.
Thus each constant $K_{i}$ must be positive; if not, the undistorted
nematic conformation would not correspond to a minimum of the free 
energy $F_{d}$. In
\cite{[1997_Kapanowski]} 
18 basic deformations proper for
the continuum theory of biaxial nematics were given. They were divided
into five groups and connected with relevant elastic constants:
3 twists (for $K_{iiii}$), 
6 splays and bends (for $K_{ijij}$),
3 modified twists (for $L_{ijk}$) and two groups of 3 double twists
(for $K_{iijj}$ and for $K_{ijji}$).
Inside the formulas for deformations a parameter $q$
was used ($1/q$ was a certain length). Small $q$
meant a small deformation and a conformation close to the
uniform one $(\vec{L}^{(0)},\vec{M}^{(0)},\vec{N}^{(0)})$. 
The vectors of the local frame were expanded into 
a power series with respect to $q$:
\begin{eqnarray}
\vec{L} & = & \vec{L}^{(0)}+q\vec{L}^{(1)}
+q^{2}\vec{L}^{(2)}+\ldots,
\nonumber \\
\vec{M} & = & \vec{M}^{(0)}+q\vec{M}^{(1)}
+q^{2}\vec{M}^{(2)}+\ldots,
\nonumber \\
\vec{N} & = & \vec{N}^{(0)}+q\vec{N}^{(1)}
+q^{2}\vec{N}^{(2)}+\ldots\ .
\label{eq:L=L0+eL1}
\end{eqnarray}
It appeared that the most important in Eq. (\ref{eq:L=L0+eL1})
were terms linear in $q$.
They were sufficient to calculate the distortion free energy up to the
second order in $q$, and to calculate the elastic constants
of biaxial nematic liquid crystals.
We add that for $K_{ii}$ one should use the same deformations as for
$K_{iiii}$ because the relevant terms have different dependence
on the parameter $q$ and they do not mix.

\section{Microscopic approach}
\label{sec:micro}

In this section we focus on the microscopic analysis 
of cholesteric liquid crystals.
Let us consider a dilute gas of $N$ molecules contained in a volume $V$,
at the temperature $T$ (in Kelvine). Let it be a binary mixture
of $N_A$ molecules $A$ and $N_B$ molecules $B$, 
where $N=N_A+N_B$.
We assume that molecules are rigid blocks
with three translational and three rotational degrees of freedom.
The state of a molecule $i$ is described by 
a vector of the position $\vec{r}_{i}$ 
and the orientation $R_{i}=(\phi_{i},\theta_{i},\psi_{i})$,
where $\phi_{i}$, $\theta_{i}$ and $\psi_{i}$ 
are the three Euler angles.
On the other hand, one can use the set of the three orthonormal
vectors $(\vec{l},\vec{m},\vec{n})$.
In a space-fixed reference frame 
$(\vec{e}_{x},\vec{e}_{y},\vec{e}_{z})$
we can express them as
\begin{equation}
\vec{l} = l_{\alpha}\vec{e}_{\alpha},
\ \vec{m} = m_{\alpha}\vec{e}_{\alpha},
\ \vec{n} = n_{\alpha}\vec{e}_{\alpha}.
\end{equation}
Let $m_I$, $J_{Ix}$, $J_{Iy}$ and $J_{Iz}$ 
denote the mass of a molecule $I$ ($I=A$ or $B$) and
the three moments of inertia, respectively.
Apart from that, we denote
$\vec{u}=\vec{r}_{2}-\vec{r}_{1}=u\vec{\Delta}$,
$R_{u}=(\phi,\theta,\psi)$
are the three Euler angles for $\vec{\Delta}$. 
In fact, $\psi$ is not used, because
$\vec{\Delta}=(\sin\theta\cos\phi,\sin\theta\sin\phi,\cos\theta)$ 
and $R_{u}$ will appear inside the rotation matrix elements 
$D^{(l)}_{\mu0}$.

We assume that molecules interact via two-body short-range forces
which depend on the distance between molecules and their 
orientations; $\Phi^{IJ}_{12}(u,R_{u},R_{1},R_{2})$
gives the potential energy of interactions ($I,J=A,B$).
Now we would like to discuss in detail symmetries of the energy 
$\Phi^{IJ}_{12}$ and molecules, because it is crucial for the 
forming of the cholesteric phase.
We note that the Euler angles always enter any formulas via 
the standard rotation matrix elements $D^{(l)}_{\mu\nu}$.
There is a problem how to incorporate the inversion $C_{i}$
into this formalism because the inversion can not be expressed 
by means of rotations.
However, one can try to simulate the inversion.
We start from the vector $\vec{\Delta}$.
After the inversion operation it gives $-\vec{\Delta}$.
Let us denote the Euler angles after the inversion by ${\bf I}R_{u}$.
\begin{equation}
\label{eq:inversion1}
D^{(l)}_{\mu0}({\bf I}R_{u})
=D^{(l)}_{\mu0}(\phi+\pi,\pi-\theta,\Psi(\psi))
=(-1)^l D^{(l)}_{\mu0}(R_{u}).
\end{equation}
We used an unknown function $\Psi$ of the angle $\psi$
because the third Euler angle is not determined before and after
the inversion. Note that the relation (\ref{eq:inversion1})
is often generalized to the form 
\cite{[1972_Blum_Torruella]}
\begin{equation}
\label{eq:inversion2}
D^{(l)}_{\mu\nu}({\bf I}R_{u})=(-1)^l D^{(l)}_{\mu\nu}(R_{u}).
\end{equation}
The relation (\ref{eq:inversion2}) in general is not true because
it cancels the matrix elements $D^{(l)}_{\mu\nu}$ with $l$ odd.
On the other hand such elements have to be used as described in
\cite{[1997_Kapanowski]}.

Let us rewrite the relation (\ref{eq:inversion1}) in the form
\begin{equation}
\label{eq:inversion3}
D^{(l)}_{\mu0}({\bf I}R_{u})=(-1)^l D^{(l)}_{\mu0}(R_{u})=
D^{(l)}_{\mu0}(R_{u}R_{z}(\pi)).
\end{equation}
Thus, for the matrix elements $D^{(l)}_{\mu0}$ the inversion can be
connected with the rotation $R_{z}(\pi)=(0,\pi,0)$ and the $C_{2}$
symmetry group. This means that there is no difference between
$D_{\infty}$ and $D_{\infty h}$ within the considered formalism.

We postulate that the invariance with respect to the inversion means
the invariance with respect to the three separate rotations
$R_{x}(\pi)$, $R_{y}(\pi)$ and $R_{z}(\pi)$ and one should replace
$D^{(l)}_{\mu\nu}({\bf I}R_{u})$ with
$D^{(l)}_{\mu\nu}(R_{u}R_{x}(\pi))$,
$D^{(l)}_{\mu\nu}(R_{u}R_{y}(\pi))$ and
$D^{(l)}_{\mu\nu}(R_{u}R_{z}(\pi))$, respectively.
This connects $C_{i}$ with the $D_{2}$ symmetry group .
This means that there is no difference between $D_{2}$ and $D_{2h}$
within the considered formalism.
It is known that $R_{x}(\pi)=R_{y}(\pi)R_{z}(\pi)$ so only two rotations
have to be used.

Now we are in a position to discuss all symmetries 
of the potential energy $\Phi^{IJ}_{12}$.
\begin{enumerate}
\item
{\em Translational invariance:}
It is satisfied because $\Phi_{12}^{IJ}$ depends on $\vec{u}$.
\item
{\em Rotational invariance:}
$\Phi_{12}^{IJ}$ should not depend on a choice of a reference frame. 
It means that for any rotation $R$
\begin{equation}
\Phi_{12}^{IJ}(u,RR_{u},RR_{1},RR_{2}) = 
\Phi_{12}^{IJ}(u,R_{u},R_{1},R_{2}).
\end{equation}
We add that generally it does not have to be satisfied because 
the interactions between two molecules can be modified by the
presence of others molecules, especially in an ordered phase.
\item
{\em Invariance with respect to the permutation 
of identical molecules:}
\begin{equation}
\Phi^{II}_{12}(u,{\bf I}R_{u},R_{2},R_{1}) =
\Phi^{II}_{12}(u,R_{u},R_{1},R_{2}).
\end{equation}
\item
{\em Invariance with respect to the symmetry operations of molecules:}
for biaxial molecules we apply operations 
from the $D_{2 h}$ symmetry group.
For the molecule $I$
\begin{equation}
\Phi_{12}^{IJ}(u,R_{u},R_{1}R_{z}(\pi),R_{2}) =
\Phi_{12}^{IJ}(u,R_{u},R_{1},R_{2}),
\end{equation}
\begin{equation}
\Phi_{12}^{IJ}(u,R_{u},R_{1}R_{y}(\pi),R_{2}) =
\Phi_{12}^{IJ}(u,R_{u},R_{1},R_{2}),
\end{equation}
and for the molecule $J$
\begin{equation}
\Phi_{12}^{IJ}(u,R_{u},R_{1},R_{2}R_{z}(\pi)) =
\Phi_{12}^{IJ}(u,R_{u},R_{1},R_{2}),
\end{equation}
\begin{equation}
\Phi_{12}^{IJ}(u,R_{u},R_{1},R_{2}R_{y}(\pi)) =
\Phi_{12}^{IJ}(u,R_{u},R_{1},R_{2}).
\end{equation}
If the molecule $I$ is uniaxial, we should add
operations from the $D_{\infty h}$ symmetry group.
For any $\alpha$
\begin{equation}
\Phi_{12}^{IJ}(u,R_{u},R_{1}R_{z}(\alpha),R_{2}) =
\Phi_{12}^{IJ}(u,R_{u},R_{1},R_{2}).
\end{equation}
\item
{\em Invariance with respect to complex conjugation:}
$\Phi_{12}^{IJ}$ should be a real function.
\item
{\em Chirality of the interactions:} the interactions between
molecules are {\em nonchiral} when
\begin{equation}
\Phi_{12}^{IJ}(u,{\bf I}R_{u},{\bf I}R_{1},{\bf I}R_{2}) =
\Phi_{12}^{IJ}(u,R_{u},R_{1},R_{2}),
\end{equation}
where ${\bf I}R$ denotes the Euler angles $R$ after the inversion.
In the opposite case the interactions are chiral and this leads to 
the forming of the cholesteric phase. 
\end{enumerate}

The microscopic free energy of the binary mixture has the form 
\cite{[1995_Chrzanowska_Sokalski]}
\begin{eqnarray}
\label{eq:betaFmix}
\beta F & = & \sum_{I=A,B} \int  {d(1)} G_{I}(1)
 \left\{ \ln [ G_{I}(1) \Lambda_{I} ]-1 \right\}
\nonumber \\
& &- {\frac {1}{2}} \sum_{I,J=A,B} \int  {d(1)}{d(2)}
 G_{I}(1)  G_{J}(2)  f_{12}^{IJ} ,
\end{eqnarray}
where $G_{I}(1)=G_{I}(\vec{r}_{1},R_{1})$
are the one-particle distribution functions with the normalizations
\begin{equation}
\label{eq:normaGI}
\int  {d(1)}   G_{I}(1) = N_{I},
\end{equation}
$d(1) = d\vec{r}_{1} dR_{1}$, 
$f_{12}^{IJ}=\exp(-\beta\Phi_{12}^{IJ})-1$ are the Mayer functions,
$\beta = 1/k_{B} T$, and
\begin{equation}
\Lambda_{I} =
\left( {\frac {h^{2} \beta}{2 \pi}} \right)^{3}
\left( m_{I}^{3} J_{Ix} J_{Iy} J_{Iz} \right)^{-1/2}.
\end{equation}
Our set of state variables consists of $T$, $V$, $N_{A}$, and $N_{B}$.
The free energy (\ref{eq:betaFmix}) consists of the ideal terms (with
$\Lambda_{I}$) and the excess terms directly related to intermolecular 
forces. The ideal terms are that of the ideal gas.

The expression (\ref{eq:betaFmix}) was derived systematically 
for binary mixtures from the Bogoliubov-Born-Green-Kirkwood-Yvon 
hierarchy equations in the thermodynamic limit 
$(N\rightarrow\infty, V\rightarrow\infty, N/V=\mbox{const})$  
\cite{[1995_Chrzanowska_Sokalski]}. 
The two-particle distribution functions were expressed in terms 
of the one-particle distribution functions 
and the two-particle correlation functions
of the simple form $\exp (-\beta \Phi_{12}^{IJ})$. 
This assumption guarantees the proper limit of the unary system.

The equilibrium distributions $G_{I}$ minimizing the free energy
(\ref{eq:betaFmix}) satisfy
\begin{equation}
\label{eq:lnG-sum_int}
\ln[G_{I}(1)\Lambda_{I}] - \sum_{J=A,B} 
\int  {d(2)}  G_{J}(2)  f_{12}^{IJ} = \mbox{const} .
\end{equation}
In the homogeneous phase the distribution function $G_{I}$ 
do not depend on the position of a molecule and 
$G_{I}(1)=G_{0I}(R_{1})$.
In order to obtain $G_{0I}$ one should solve the equations
(\ref{eq:lnG-sum_int}) together with (\ref{eq:normaGI}).

In order to define the microscopic distortion free-energy density 
$f_{d}$ one should also identify the homogeneous free-energy density 
$f_{0}$. 
Note that the terms with $\Lambda_{I}$ in (\ref{eq:betaFmix}) 
are local: the integrands involve the distribution function 
for a single point only.
Other terms couples the distributions at neighbouring points.
The dependence of the free energy on the spatial variations 
of the ordering will be found by expanding this terms 
in the gradients of $G_{J}$
\cite{[1976_Straley]}. 
Substituting the Taylor expansion of $G_{J}(\vec{r}_{1}+\vec{u})$
into (\ref{eq:betaFmix}) we get
\begin{equation}
F = \int  {d\vec{r}} [f_{0}(\vec{r})+f_{d}(\vec{r})],
\end{equation}
where
\begin{eqnarray}
\beta f_{0} (\vec{r}) & = & \sum_{I=A,B} 
\int  {dR} G_{I}(\vec{r},R) 
\left\{ \ln [ G_{I}(\vec{r},R) \Lambda_{I} ]-1 \right\}
\nonumber\\
& & \mbox{} - {\frac {1}{2}} \sum_{I,J=A,B} 
\int  {dR_{1}}{dR_{2}} {d\vec{u}}
G_{I}(\vec{r},R_{1}) G_{J}(\vec{r},R_{2}) f_{12}^{IJ} ,
\label{def:f0}
\\
\beta f_{d}(\vec{r}) & = &
- {\frac {1}{2}} \sum_{I,J=A,B} 
\int  {dR_{1}}{dR_{2}} {d\vec{u}} G_{I}(\vec{r},R_{1}) 
[(\vec{u} \cdot \nabla)G_{J}(\vec{r},R_{2})] f_{12}^{IJ}
\nonumber\\
& & \mbox{} + {\frac {1}{4}} \sum_{I,J=A,B} 
\int  {dR_{1}}{dR_{2}} {d\vec{u}}
[(\vec{u}\cdot\nabla)G_{I}(\vec{r},R_{1})] 
\nonumber\\
& & \times [(\vec{u}\cdot\nabla)G_{J}(\vec{r},R_{2})] f_{12}^{IJ}.
\label{def:fd}
\end{eqnarray}
Note that an integration by parts has been used to combine 
the second order terms and the surface terms have been neglected 
thanks to the thermodynamic limit. 
The definition (\ref{def:f0}) is equivalent
to that by Poniewierski and Stecki 
\cite{[1979_Poniewierski_Stecki]}.
This is a well-founded assumption if we also assume 
slow variations of the vectors $(\vec{L},\vec{M},\vec{N})$. 
We will also restrict the one-particle distribution function 
$G_{I}$ to the class of $G_{0I}$ functions. 
This method was succesfully used in the past 
\cite{[1973_Straley],[1991_Marrucci_Greco]}.
As we expect, for the homogeneous phase $f_{d}$ becomes equal to zero. 

It was shown in
\cite{[1995_FKS],[1997_Kapanowski]}
that in the case of the homogeneous biaxial nematic phase composed 
of biaxial molecules, the one-particle distribution function $G_{0I}$
depends on four arguments:
\begin{equation}
G_{0I}(R)=G_{0I}(
\vec{l}\cdot\vec{e}_{x},
\vec{l}\cdot\vec{e}_{z},
\vec{n}\cdot\vec{e}_{x},
\vec{n}\cdot\vec{e}_{z}).
\end{equation}
We postulate that the distribution of the distorted phase 
$G_{I}(\vec{r},R)$ can be written as
\begin{equation}
G_{I}(\vec{r},R)=G_{0I}(
\vec{l}\cdot\vec{L}(\vec{r}),
\vec{l}\cdot\vec{N}(\vec{r}),
\vec{n}\cdot\vec{L}(\vec{r}),
\vec{n}\cdot\vec{N}(\vec{r})),
\end{equation}
where the reference frame 
$(\vec{e}_{x},\vec{e}_{y},\vec{e}_{z})$
is replaced with the local frame 
$(\vec{L}(\vec{r}),\vec{M}(\vec{r}),\vec{N}(\vec{r}))$. 

\section{Elastic constants}
\label{sec:elastic}

Now we are in a position to substitute the basic deformations 
into the microscopic distortion free-energy density (\ref{def:fd}) 
and to the phenomenological distortion free-energy density 
(\ref{eq:d2}) and (\ref{eq:d8}). 
As a result of the comparison we get 
the microscopic formulas for the elastic constants. 
To make them more compact we write
\begin{eqnarray}
U_{\alpha}^{I} & = &
\partial_{1}G_{0I} l_{\alpha}+\partial_{3}G_{0I} n_{\alpha},
\nonumber \\
W_{\alpha}^{I} & = &
\partial_{2}G_{0I} l_{\alpha}+\partial_{4}G_{0I} n_{\alpha}.
\label{def:UaWa}
\end{eqnarray}
Microscopic expressions for the chiral elastic constants 
of the biaxial cho\-les\-teric phase are as follows:
\begin{eqnarray}
\beta K_{11} & = & - {\frac{1}{2}} \int  {dR_{1}}{dR_{2}}{d\vec{u}} 
u_{x} \sum_{I,J=A,B} f_{12}^{IJ} G_{0I}(R_{1}) W_{2y}^{J},
\nonumber \\
\beta K_{22} & = & - {\frac{1}{2}} \int  {dR_{1}}{dR_{2}}{d\vec{u}} 
u_{y}\sum_{I,J=A,B} f_{12}^{IJ} G_{0I}(R_{1}) (U_{2z}^{J}-W_{2x}^{J}),
\nonumber \\
\beta K_{33} & = & {\frac{1}{2}} \int  {dR_{1}}{dR_{2}}{d\vec{u}} 
u_{z}\sum_{I,J=A,B} f_{12}^{IJ} G_{0I}(R_{1}) U_{2y}^{J}.
\label{eq:Kii}
\end{eqnarray}
Microscopic expressions for the nonchiral elastic constants are 
the following.
            The first group is,
\begin{eqnarray}
\label{eq:K1111endmix}
\beta K_{1111} & = & {\frac{1}{2}} \int  {dR_{1}}{dR_{2}}{d\vec{u}} 
u_{x}^{2} \sum_{I,J=A,B} f_{12}^{IJ}  W_{1y}^{I} W_{2y}^{J}, 
\\
\beta K_{2222} & = & {\frac{1}{2}} \int  {dR_{1}}{dR_{2}}{d\vec{u}} 
u_{y}^{2} \sum_{I,J=A,B} f_{12}^{IJ}  
(U_{1z}^{I}-W_{1x}^{I}) (U_{2z}^{J}-W_{2x}^{J}), 
\\
\beta K_{3333} & = & {\frac{1}{2}} \int  {dR_{1}}{dR_{2}}{d\vec{u}} 
u_{z}^{2} \sum_{I,J=A,B} f_{12}^{IJ}  U_{1y}^{I} U_{2y}^{J}.
\end{eqnarray}
           The second group is,
\begin{eqnarray}
\beta K_{1212} & = & {\frac{1}{2}} \int  {dR_{1}}{dR_{2}}{d\vec{u}} 
u_{x}^{2} \sum_{I,J=A,B}
f_{12}^{IJ}  (U_{1z}^{I}-W_{1x}^{I}) (U_{2z}^{J}-W_{2x}^{J}), 
\\
\beta K_{1313} & = & {\frac{1}{2}} \int  {dR_{1}}{dR_{2}}{d\vec{u}} 
u_{x}^{2} \sum_{I,J=A,B} f_{12}^{IJ}  U_{1y}^{I} U_{2y}^{J}, 
\\
\beta K_{2121} & = & {\frac{1}{2}} \int  {dR_{1}}{dR_{2}}{d\vec{u}} 
u_{y}^{2} \sum_{I,J=A,B} f_{12}^{IJ}  W_{1y}^{I} W_{2y}^{J},   
\\
\beta K_{2323} & = & {\frac{1}{2}} \int  {dR_{1}}{dR_{2}}{d\vec{u}} 
u_{y}^{2} \sum_{I,J=A,B} f_{12}^{IJ}  U_{1y}^{I} U_{2y}^{J}, 
\\
\beta K_{3131} & = & {\frac{1}{2}} \int  {dR_{1}}{dR_{2}}{d\vec{u}} 
u_{z}^{2} \sum_{I,J=A,B} f_{12}^{IJ}  W_{1y}^{I} W_{2y}^{J}, 
\\
\beta K_{3232} & = & {\frac{1}{2}} \int  {dR_{1}}{dR_{2}}{d\vec{u}} 
u_{z}^{2} \sum_{I,J=A,B}
f_{12}^{IJ} (U_{1z}^{I}-W_{1x}^{I}) (U_{2z}^{J}-W_{2x}^{J}).
\end{eqnarray}
              The third group is,
\begin{equation}
\label{eq:Lijk=0}
L_{123} = L_{231} = L_{312} =0.
\end{equation}
               The fourth group is,
\begin{eqnarray}
\beta K_{1122} & = & {\frac{1}{4}} \int  {dR_{1}}{dR_{2}}{d\vec{u}}
u_{x} u_{y} 
\nonumber \\
& & \times \sum_{I,J=A,B} f_{12}^{IJ} 
[(U_{1z}^{I}-W_{1x}^{I})W_{2y}^{J}+W_{1y}^{I}(U_{2z}^{J}-W_{2x}^{J})],
\\
\beta K_{2233}&  = & {\frac{1}{4}} \int  {dR_{1}}{dR_{2}}{d\vec{u}}
u_{y}u_{z} 
\nonumber \\
& & \times \sum_{I,J=A,B} f_{12}^{IJ}
[-U_{1y}^{I}(U_{2z}^{J}-W_{2x}^{J})-(U_{1z}^{I}-W_{1x}^{I}) U_{2y}^{J}],
\\
\beta K_{1133} & = & {\frac{1}{4}} \int  {dR_{1}}{dR_{2}}{d\vec{u}}
u_{x}u_{z}\sum_{I,J=A,B} f_{12}^{IJ}
[-U_{1y}^{I} W_{2y}^{J}-W_{1y}^{I}U_{2y}^{J}].
\end{eqnarray}
             The fifth group is,
\begin{equation}
\label{eq:Kijji=Kiijj}
K_{1221}=K_{1122},\  K_{1331}=K_{1133},\  K_{2332}=K_{2233}.
\label{eq:K2332endmix}
\end{equation}

In the case of the uniaxial cholesteric phase the expression 
for the chiral elastic constant is
\begin{equation}
\beta K_{0}  =  {\frac{1}{2}} \int  {dR_{1}}{dR_{2}}{d\vec{u}} 
u_{y} \sum_{I,J=A,B} f_{12}^{IJ} G_{0I}(R_{1}) W_{2x}^{J}.
\end{equation}
The expressions for the nonchiral elastic constants are as follows:
\begin{eqnarray}
\label{eq:K1mix}
\beta K_{1}& = & {\frac {1}{2}} \int  {dR_{1}}{dR_{2}}{d\vec{u}}
 u_{x}^{2}    \sum_{I,J=A,B} f_{12}^{IJ} W_{1x}^{I} W_{2x}^{J},  
\\
\label{eq:K2mix}
\beta K_{2} & = &{\frac {1}{2}} \int  {dR_{1}}{dR_{2}}{d\vec{u}}
 u_{y}^{2}  \sum_{I,J=A,B} f_{12}^{IJ} W_{1x}^{I} W_{2x}^{J},  
\\
\label{eq:K3mix}
\beta K_{3} & = & {\frac {1}{2}} \int  {dR_{1}}{dR_{2}}{d\vec{u}}
 u_{z}^{2}\sum_{I,J=A,B} f_{12}^{IJ} W_{1x}^{I} W_{2x}^{J},  
\\
\label{eq:K4mix}
K_{4} & = & {\frac {1}{2}} (K_{1} + K_{2}), 
\\  
\label{eq:K5mix}
K_{5} & = & 0.
\end{eqnarray}
Let us show how the elastic constants determine the state of cholesteric
liquid crystals. It is very important that, when we look for the physical
state of the liquid crystal in the thermodynamic limit, we should remove
all surface terms from the phenomenological free-energy densities.
The reason is that in the thermodynamic limit all surface terms are
negligible. We used the surface terms to derive the expressions
for the elastic constants to assure the consistency because the
surface terms are present in the hidden form in the microscopic
free-energy density.
The equilibrium distortion of the uniaxial cholesteric phase 
is a pure twist
\begin{equation}
\vec{N}(\vec{r})=[0,-\sin(q x),\cos(q x)]
\end{equation}
where $q=K_{0}/K_{2}$, $P=2\pi/|q|$ is the cholesteric pitch
and $f_{d}=-K_{0}^{2}/2K_{2}$. Note that the sign of $K_{0}$ 
distinguishes between right- and left-handed helices.

For the biaxial cholesteric liquid crystals the equilibrium distortion 
is composed of three twists with respect to the orthogonal axes.
In the limit of small distortions the vectors of the local frame have 
the form
\begin{eqnarray}
\vec{L} & = & (1,q_{3}z,-q_{2}y),
\nonumber \\
\vec{M} & = & (-q_{3}z,1,q_{1}x),
\nonumber \\
\vec{N} & = & (q_{2}y,-q_{1}x,1),
\end{eqnarray}
where $q_{i}=W_{i}/W$,
\begin{eqnarray}
W & = & K_{1111}K_{2222}K_{3333}+2 K_{1122}K_{1133}K_{2233} 
\nonumber\\
& & \mbox{} - K_{1111}K_{2233}^{2}-K_{2222}K_{1133}^{2}-K_{3333}K_{1122}^{2}, 
\nonumber\\
W_{1} & = & K_{11} (K_{2222}K_{3333}-K_{2233}^{2})
\nonumber\\
& & \mbox{} + K_{22} (K_{1133}K_{2233}-K_{1122}K_{3333})
\nonumber\\
& & \mbox{} + K_{33} (K_{1122}K_{2233}-K_{1133}K_{2222}),
\nonumber\\
W_{2} & = & K_{11} (K_{1133}K_{2233}-K_{1122}K_{3333})
\nonumber\\
& & \mbox{} + K_{22} (K_{1111}K_{3333}-K_{1133}^{2})
\nonumber\\
& & \mbox{} + K_{33} (K_{1122}K_{1133}-K_{1111}K_{2233}),
\nonumber\\
W_{3} & = & K_{11} (K_{1122}K_{2233}-K_{1133}K_{2222})
\nonumber\\
& & \mbox{} + K_{22} (K_{1133}K_{1122}-K_{1111}K_{2233})
\nonumber\\
& & \mbox{} + K_{33} (K_{1111}K_{2222}-K_{1122}^{2}).
\end{eqnarray}
The distortion free-energy density at the minimum is
\begin{eqnarray}
f_{d} & = & [2 K_{11}K_{22}(K_{3333}K_{1122}-K_{1133}K_{2233})
\nonumber\\
& & \mbox{} + 2 K_{11}K_{33}(K_{2222}K_{1133}-K_{1122}K_{2233})
\nonumber\\
& & \mbox{} + 2 K_{22}K_{33}(K_{1111}K_{2233}-K_{1122}K_{1133})
\nonumber\\
& & \mbox{} + K_{11}^{2}(K_{2233}^{2}-K_{2222}K_{3333})
\nonumber\\
& & \mbox{} + K_{22}^{2}(K_{1133}^{2}-K_{1111}K_{3333})
\nonumber\\
& & \mbox{} + K_{33}^{2}(K_{1122}^{2}-K_{1111}K_{2222})]/2W^{2}.
\end{eqnarray}
It is important that a global minimum of $f_{d}$  exists only if the matrix
\begin{equation}
\label{macierzK}
\left[
\begin{array}{ccc}
K_{1111} & K_{1122} & K_{1133} \\
K_{1122} & K_{2222} & K_{2233} \\
K_{1133} & K_{2233} & K_{3333} \\
\end{array}
\right] 
\end{equation}
is positive definite. In the opposite case there is no stable 
biaxial cholesteric (and nematic) phase. 
Taking the uniaxial phase limit, we get simpler conditions. 
We can say that there is a stable cholesteric or nematic phase only if
\begin{equation}
\label{eq:k1<3k2}
K_{1} < 3K_{2}.
\end{equation}
As far as we know almost all theories and all real and computer 
experiments are in the agreement with the relation (\ref{eq:k1<3k2}).
The equality $K_{1}=3 K_{2}$ appears in some theories with hard 
molecules 
\cite{[1973_Straley],[1982_Kimura]}.

\section{Exemplary calculations}
\label{sec:corner}

The aim of this section is to express the elastic constants by means of
the order parameters which can be measured in experiments.
We will apply the Corner potential energy of the interactions because 
in principle it allows detailed calculations without any additional 
approximations. On the other hand, it is quite realistic.
The Corner potential energy has the form
$\Phi_{12}^{IJ}(u/\sigma^{IJ})$,
where $\sigma^{IJ}$ depends on orientations $R_{1}$, $R_{2}$ and $R_{u}$.
For $\sigma^{IJ}$ one can write the general expansion proposed by
Blum and Torruela 
\cite{[1972_Blum_Torruella]}. 
It involvs the 3-j Wigner symbols and the standard rotation matrix 
elements. The same expression was used to describe the interactions 
of biaxial molecules in
\cite{[1995_FKS]}. 
In the case of biaxial molecules the lowest 
order terms of the expansion give
\begin{eqnarray}
\label{def:sigma}
\sigma^{IJ} & = & \sigma_{0}^{IJ}
+ \sigma_{11}^{IJ} ( \vec{\Delta} \cdot \vec{n}_{1} )^{2}
+ \sigma_{12}^{IJ} ( \vec{\Delta} \cdot \vec{n}_{2} )^{2}
+ \sigma_{2}^{IJ} ( \vec{n}_{1} \cdot \vec{n}_{2} )^{2}
\nonumber \\
& & \mbox{} + \sigma_{31}^{IJ} ( \vec{\Delta} \cdot \vec{l}_{1} )^{2}
+ \sigma_{32}^{IJ} ( \vec{\Delta} \cdot \vec{l}_{2} )^{2}
+ \sigma_{4}^{IJ} ( \vec{l}_{1} \cdot \vec{l}_{2} )^{2}
\nonumber \\
& & \mbox{} + \sigma_{51}^{IJ} 
( \vec{l}_{1} \cdot \vec{n}_{2} )^{2}
+ \sigma_{52}^{IJ} ( \vec{n}_{1} \cdot \vec{l}_{2} )^{2}
 + \sigma_{6}^{IJ}( \vec{n}_{1} \cdot \vec{n}_{2} )
( \vec{n}_{1} \times \vec{n}_{2} )\cdot \vec{\Delta}
\nonumber \\
& & \mbox{} + \sigma_{71}^{IJ} 
[( \vec{l}_{1} \cdot \vec{n}_{2} )
( \vec{l}_{1} \times \vec{n}_{2} )\cdot \vec{\Delta}
- ( \vec{m}_{1} \cdot \vec{n}_{2} )
( \vec{m}_{1} \times \vec{n}_{2} )\cdot \vec{\Delta}]
\nonumber \\
& & \mbox{} + \sigma_{72}^{IJ} 
[( \vec{n}_{1} \cdot \vec{l}_{2} )
 ( \vec{n}_{1} \times \vec{l}_{2} )\cdot \vec{\Delta}
-( \vec{n}_{1} \cdot \vec{m}_{2} )
 ( \vec{n}_{1} \times \vec{m}_{2} )\cdot \vec{\Delta} ]
\nonumber \\
& & \mbox{} + \sigma_{8}^{IJ} [
( \vec{l}_{1} \cdot \vec{l}_{2} )
( \vec{l}_{1} \times \vec{l}_{2} )\cdot \vec{\Delta}
+( \vec{m}_{1} \cdot \vec{m}_{2} )
 ( \vec{m}_{1} \times \vec{m}_{2} )\cdot \vec{\Delta}
\nonumber \\
& & \mbox{} - ( \vec{l}_{1} \cdot \vec{m}_{2} )
( \vec{l}_{1} \times \vec{m}_{2} )\cdot \vec{\Delta}
-( \vec{m}_{1} \cdot \vec{l}_{2} )
 ( \vec{m}_{1} \times \vec{l}_{2} )\cdot \vec{\Delta} ].
\end{eqnarray}
There are 13 molecular parameters that determine the main features of
the interactions between biaxial molecules. 
We would like to add some comments to this long expression. 
The terms with $\sigma_{0}^{IJ}$, $\sigma_{11}^{IJ}$, $\sigma_{12}^{IJ}$,
and $\sigma_{2}^{IJ}$ describe the nonchiral interactions of uniaxial 
molecules. 
The term with $\sigma_{6}^{IJ}$ describes chiral part of their 
interactions. A similar term was used in the past 
\cite{[1977_Lin-Liu_Shih_Woo],[1978_Rajan_Woo]}.
The terms with $\sigma_{0}^{IJ}$, $\sigma_{11}^{IJ}$, $\sigma_{12}^{IJ},
\sigma_{2}^{IJ}$, $\sigma_{31}^{IJ}$, $\sigma_{32}^{IJ}$, 
$\sigma_{4}^{IJ}$, $\sigma_{51}^{IJ}$, and $\sigma_{52}^{IJ}$ 
describe the nonchiral interactions of biaxial molecules 
\cite{[1995_FKS],[1997_Kapanowski]}.
The additional terms with $\sigma_{6}^{IJ}$, $\sigma_{71}^{IJ}$,
$\sigma_{72}^{IJ}$, and $\sigma_{8}^{IJ}$ 
allow to decribe the chiral interactions of biaxial molecules. 
As far as we know they were not presented in the literature. 
Note that the terms with $\vec{\Delta}$
couple the translational and rotational degrees of freedom.

In order to set the proper values of the parameters $\sigma_{i}^{IJ}$
one should investigated the equipotential surfaces of the potential
energy. The excluded volume method described in
\cite{[1995_FKS]}
can be helpful. Molecules $I$ are described by 
\begin{eqnarray}
& \sigma_{0}^{II},
\ \sigma_{1}^{II} \equiv \sigma_{11}^{II} = \sigma_{12}^{II},
\ \sigma_{2}^{II},
\ \sigma_{3}^{II} \equiv \sigma_{31}^{II} = \sigma_{32}^{II},
\ \sigma_{4}^{II}, &
\nonumber \\
& \sigma_{5}^{II} \equiv \sigma_{51}^{II} = \sigma_{52}^{II},
\ \sigma_{6}^{II},
\ \sigma_{7}^{II} \equiv \sigma_{71}^{II} = \sigma_{72}^{II},
\ \sigma_{8}^{II}. &
\end{eqnarray}
The interactions between molecles $I$ and $J$ we describe by
\begin{eqnarray}
& \sigma_{0}^{IJ}=(\sigma_{0}^{II}+\sigma_{0}^{JJ})/2,
\ \sigma_{11}^{IJ}=\sigma_{1}^{II},
\ \sigma_{12}^{IJ}=\sigma_{1}^{JJ},&
\nonumber \\
& \sigma_{2}^{IJ}=(\sigma_{2}^{II}+\sigma_{2}^{JJ})/2,
\ \sigma_{31}^{IJ}=\sigma_{3}^{II},
\ \sigma_{32}^{IJ}=\sigma_{3}^{JJ},&
\nonumber \\
& \sigma_{4}^{IJ}=(\sigma_{4}^{II}+\sigma_{4}^{JJ})/2,
\ \sigma_{51}^{IJ}=\sigma_{52}^{IJ}=(\sigma_{5}^{II}+\sigma_{5}^{JJ})/2,&
\nonumber \\
& \sigma_{6}^{IJ}=(\sigma_{6}^{II}+\sigma_{6}^{JJ})/2,
\ \sigma_{71}^{IJ}=\sigma_{72}^{IJ}=(\sigma_{7}^{II}+\sigma_{7}^{JJ})/2,&
\nonumber \\
& \sigma_{8}^{IJ}=(\sigma_{8}^{II}+\sigma_{8}^{JJ})/2.&
\end{eqnarray}
The parameters $\sigma_{i}^{IJ}$ are the molecular constants 
that in our model determine the ordering of the phase, 
the value of the elastic constants and the twist of the phase. 
As far as the functional dependence of the 
potential energy on the $u/\sigma^{IJ}$ is concerned, we have many 
possibilities and we will mention two of them, together with
a function $B_{s}(T)$ defined as
\begin{equation}
B_{s}(T)
  = \int_{0}^{\infty}  {dx} x^{s} f_{12}^{IJ}(x)
= \int_{0}^{\infty}  {dx} x^{s} [\exp (- \beta \Phi_{12}^{IJ}(x))-1].
\end{equation}
Let $\epsilon$ denotes a depth of the potential energy (we assume for 
simplicity that it is the same for both types of molecules).
\begin{enumerate}
\item
The {\em soft-core potential energy}
\begin{equation}
\Phi_{12}^{IJ}(u/ \sigma^{IJ}) =  \epsilon (\sigma^{IJ}/u)^{m},
\end{equation}
\begin{equation}
B_{s}(T) = {\frac {-1}{s+1}}
\Gamma \left( {\frac {m-s-1}{m}} \right)
\left( {\frac {\epsilon}{k_{B}T}} \right)^{(s+1)/m}.
\end{equation}
\item
The {\em square-well potential energy}
\begin{equation}
\Phi_{12}^{IJ}(u/ \sigma^{IJ}) =  \left\{
\begin{array}{ll}
+\infty   & \mbox{for} \ (u/\sigma^{IJ}) < 1  \\
-\epsilon & \mbox{for} \ 1 < (u/\sigma^{IJ}) < R_{SW}  \\
0         & \mbox{for} \  (u/\sigma^{IJ}) > R_{SW} ,
\end{array}  \right.
\end{equation}
\begin{equation}
B_{s}(T) = {\frac {1}{s+1}} 
\left[ \left( \exp (\epsilon/k_{B}T)-1 \right) (R_{SW}^{s+1}-1) -1
\right].
\end{equation}
\end{enumerate}
Let us define dimensionless functions $f_I(R)=G_{0I}(R)V/N_I$ 
with the normalization
\begin{equation}
\int  {dR}  f_I(R)=1.
\end{equation}
In order to describe systems with biaxial symmetry it is convenient to
introduce the complete set of basic functions (invariants) proper for 
that kind of symmetry. They are closely connected with the rotation 
matrix elements and have the form
\begin{equation}
F_{\mu\nu}^{(j)}(R) = \left( {\frac{1}{\sqrt{2}}} 
\right)^{2+\delta_{\mu 0}+\delta_{0\nu}}
\sum_{\rho,\sigma=\pm1} (-1)^{j(\sigma-\rho)/2}
D_{\rho\cdot\mu,\sigma\cdot\nu}^{(j)}(R),
\end{equation}
where $j$ is a nonnegative integer, $\mu,\nu$ are even. 
If $j$ is even, then $0\leq\mu\leq j$ and $0\leq\nu\leq j$.
If $j$ is odd, then $2\leq\mu\leq j$ and $2\leq\nu\leq j$.
The invariants are real and orthogonal functions
\begin{equation}
\int  {dR} F_{\mu\nu}^{(j)}(R) F_{\rho\sigma}^{(k)}(R)=
\delta_{jk} \delta_{\mu\rho} \delta_{\nu\sigma} 8\pi^{2}/(2j+1).
\end{equation}
The properties of the invariants were described in 
\cite{[1997_Kapanowski]}.
It was shown that all $F_{\mu\nu}^{(j)}$ can be expressed by means of
$(\vec{l}\cdot\vec{e}_{x})^2$,
$(\vec{l}\cdot\vec{e}_{z})^2$,
$(\vec{n}\cdot\vec{e}_{x})^2$, and
$(\vec{n}\cdot\vec{e}_{z})^2$. 
Let us show the most important invariants with $j=2$:
\begin{eqnarray}
F_{00}^{(2)} & = & [-1+3(\vec{n}\cdot\vec{e}_{z})^2]/2,
\nonumber\\
F_{02}^{(2)} & = & [-1+(\vec{n}\cdot\vec{e}_{z})^2
+2 (\vec{l}\cdot\vec{e}_{z})^2]\sqrt{3}/2,
\nonumber\\
F_{20}^{(2)} & = & [-1+(\vec{n}\cdot\vec{e}_{z})^2
+2 (\vec{n}\cdot\vec{e}_{x})^2]\sqrt{3}/2,
\nonumber\\
F_{22}^{(2)} & = & [-3+(\vec{n}\cdot\vec{e}_{z})^2
+2 (\vec{l}\cdot\vec{e}_{z})^2
+2 (\vec{n}\cdot\vec{e}_{x})^2
+4 (\vec{l}\cdot\vec{e}_{x})^2]/2.
\end{eqnarray}
For the first time all four invariants (with different normalization) 
appeared in the paper by Straley 
\cite{[1974_Straley]}.
Let us define the order parameters for our system as
\begin{equation}
\langle F_{\mu\nu}^{(j)}\rangle_I \equiv 
\int  {dR} f_I(R) F_{\mu\nu}^{(j)}(R).
\end{equation}
Note that for the completely ordered biaxial phase we have
$\langle F_{\mu\nu}^{(j)}\rangle_I =\delta_{\mu\nu}$,
thus the most important are 
$\langle F_{00}^{(2)}\rangle_I$ and
$\langle F_{22}^{(2)}\rangle_I$
($S$ and $V$ in the Straley notation).

Let us rewrite (\ref{eq:lnG-sum_int}) for the 
one-particle distribution function of the uniform phase. 
It is identical to one in the theory with nonchiral interactions. 
We define
\begin{equation}
K^{IJ}(R_{1},R_{2}) =
\int  {d\vec{\Delta}} (\sigma^{IJ}/\sigma_{0}^{IJ})^{3},
\end{equation}
\begin{equation}
\label{eq:lambda}
\lambda^{IJ} = B_{2}(T) (\sigma_{0}^{IJ})^{3}N_J/V.
\end{equation}
The kernel $K^{IJ}$ and the distribution functions $f_I$ 
can be expressed in terms of the invariants 
\begin{equation}
\label{sumK}
K^{IJ}(R,0) = \sum_j \sum_{\mu\nu}
K_{\mu\nu}^{(j)IJ} F_{\mu\nu}^{(j)}(R),
\end{equation}
\begin{equation}
\label{sumlnf}
\ln f_I(R) =  \sum_j \sum_{\mu\nu}
S_{\mu\nu}^{(j)I} F_{\mu\nu}^{(j)}(R).
\end{equation}
Note that for $\sigma^{IJ}$ given by (\ref{def:sigma})
both sums (\ref{sumK}) and (\ref{sumlnf}) are {\em finite} series.
Thus the distribution function $f_I$ is fully described 
by 35 coefficients $S_{\mu\nu}^{(j)I}$. 
We calculate the coefficients $S_{\mu\nu}^{(j)I}$ from the equations
\begin{equation}
S_{\mu\nu}^{(j)I} = \sum_{J=A,B} 
\lambda^{IJ} \sum_{\rho} K_{\rho\nu}^{(j)IJ}
\langle F_{\mu\rho}^{(j)}\rangle_J \ \mbox{for} \ j>0,
\end{equation}
\begin{equation}
\langle F_{00}^{0}\rangle_I = 1 \ \mbox{(the normalization condition)}.
\end{equation}
Now we move to the elastic constants. 
We can expand the distribution function $f_I$
in an infinite series with respect to invariants
\begin{equation}
\label{sumf}
f_I(R)=\sum_{j} \sum_{\mu\nu}\langle F_{\mu\nu}^{(j)}\rangle_I
F_{\mu\nu}^{(j)}(R) (2j+1)/(8\pi^2).
\end{equation}
All elastic constants can be written as {\em finite} sums of the form
\begin{equation}
K_{ii} = \sum \xi^{IJ}
\langle F_{\mu\nu}^{(j)} \rangle_I
\langle F_{\rho\sigma}^{(k)} \rangle_J
B_{ii,\mu\nu\rho\sigma}^{(j)(k)IJ},
\end{equation}
\begin{equation}
K_{ijkl} = \sum \eta^{IJ}
\langle F_{\mu\nu}^{(j)} \rangle_I
\langle F_{\rho\sigma}^{(k)} \rangle_J
A_{ijkl,\mu\nu\rho\sigma}^{(j)(k)IJ},
\end{equation}
where
\begin{equation}
\label{eq:ksi}
\xi^{IJ} = k_{B}T B_{3}(T)  (\sigma_{0}^{IJ})^{4} N_{I} N_{J}/V^{2},
\end{equation}
\begin{equation}
\label{eq:eta}
\eta^{IJ} = k_{B}T B_{4}(T)  (\sigma_{0}^{IJ})^{5} N_{I}N_{J}/V^{2},
\end{equation}
The coefficients 
$K_{\mu\nu}^{(j)IJ}$,
$A_{ijkl,\mu\nu\rho\sigma}^{(j)(k)IJ}$, and 
$B_{ii,\mu\nu\rho\sigma}^{(j)(k)IJ}$ 
are polynomials in $(\sigma_{i}^{IJ}/\sigma_{0}^{IJ})$ 
that can be calculated analytically. 

\subsection{Unary phase of biaxial molecules}

\begin{figure}
\begin{center}
\includegraphics[scale=0.5,angle=270]{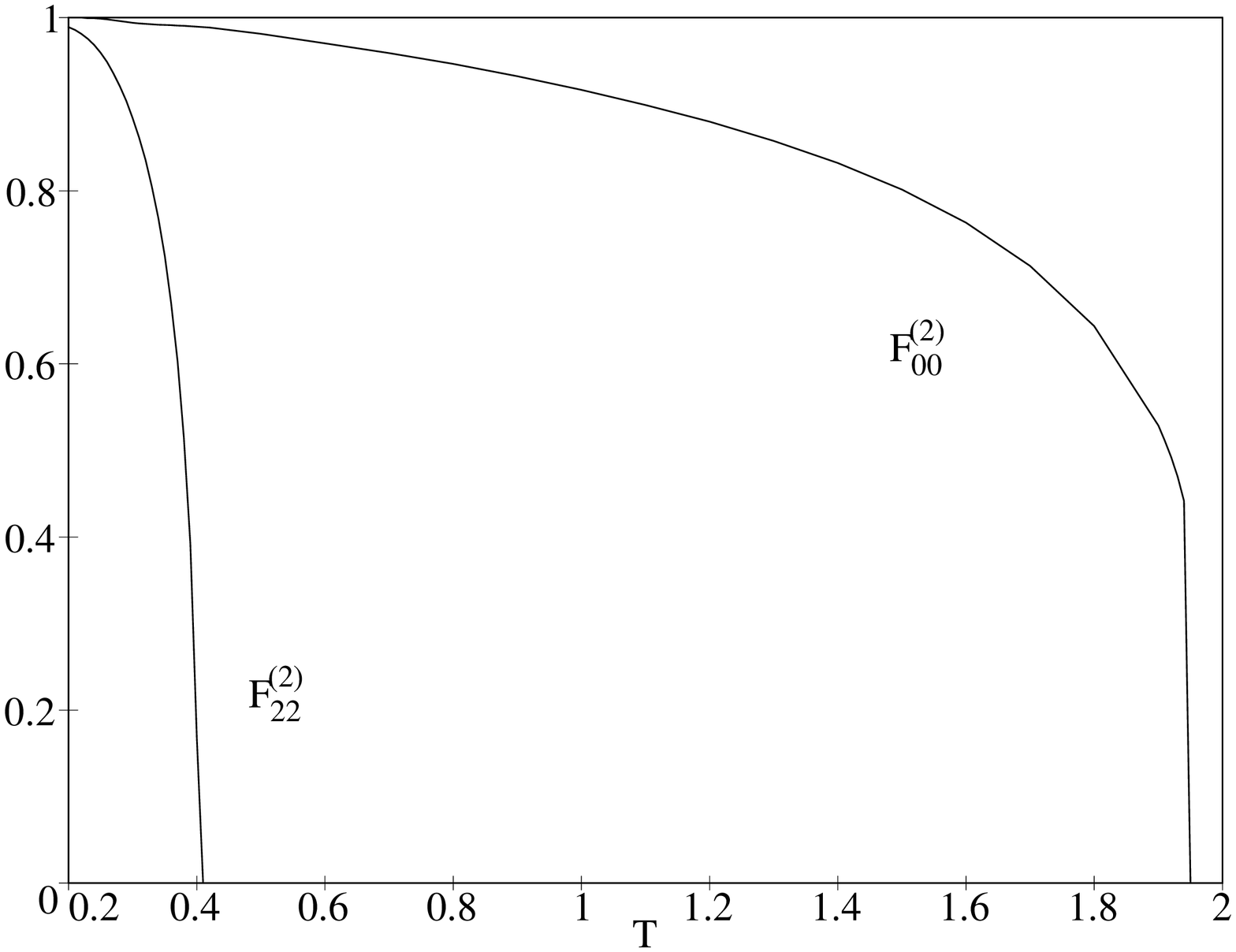}
\end{center}
\caption{
\label{fig1}
\interlinia
Temperature dependence of the order parameters
$\langle F_{00}^{(2)}\rangle$ and $\langle F_{22}^{(2)}\rangle$.}
\end{figure}

\begin{figure}
\begin{center}
\includegraphics[scale=0.5,angle=270]{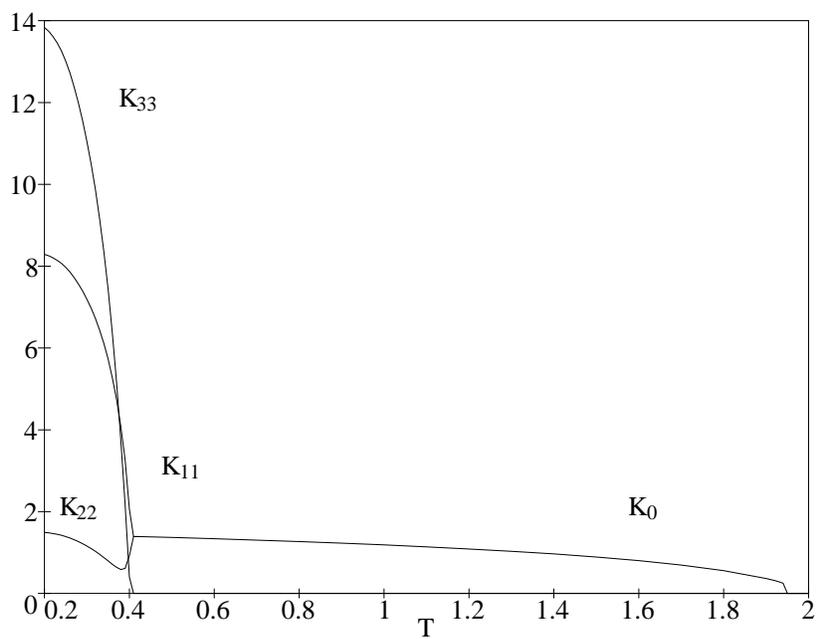}
\end{center}
\caption{
\label{fig2}
\interlinia
Temperature dependence of the chiral elastic constants $K_{ii}/\xi$.
In the uniaxial cholesteric phase (between $T=1.94$ and $T=0.41$)
we have $K_{0}=K_{11}=K_{22}$ and $K_{33}=0$.
In the biaxial cholesteric phase (below $T=0.41$) $K_{0}$ splits
into two constants and $K_{33}$ becomes nonzero.}
\end{figure}

\begin{figure}
\begin{center}
\includegraphics[scale=0.5,angle=270]{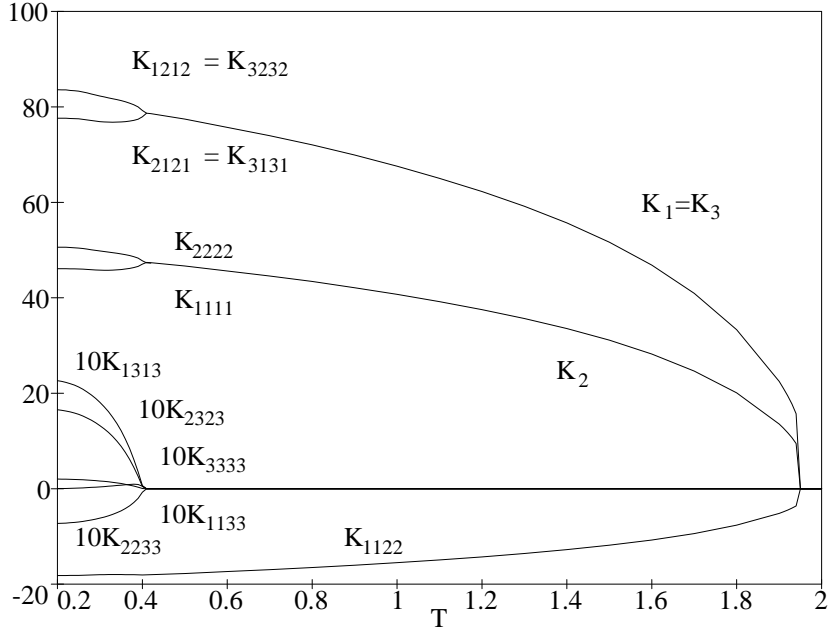}
\end{center}
\caption{
\label{fig3}
\interlinia
Temperature dependence of the nonchiral elastic constants 
$K_{ijkl}/\eta$.
In the uniaxial cholesteric phase (between $T=1.94$ and $T=0.41$)
we have $K_{1}=K_{1212}=K_{2121}$ equal to $K_{3}=K_{3131}=K_{3232}$
and $K_{2}=K_{1111}=K_{2222}$. We have also $K_{1122}=(K_{2}-K_{1})/2$.
In the biaxial cholesteric phase (below $T=0.41$)
$K_{1}$, $K_{2}$ and $K_{3}$ split into two constants,
$K_{1122}$ becomes an independent constant,
five constants become nonzero ($K_{1133}$, $K_{2233}$, $K_{1313}$,
$K_{2323}$ and $K_{3333}$).}
\end{figure}

\begin{figure}
\begin{center}
\includegraphics[scale=0.5,angle=270]{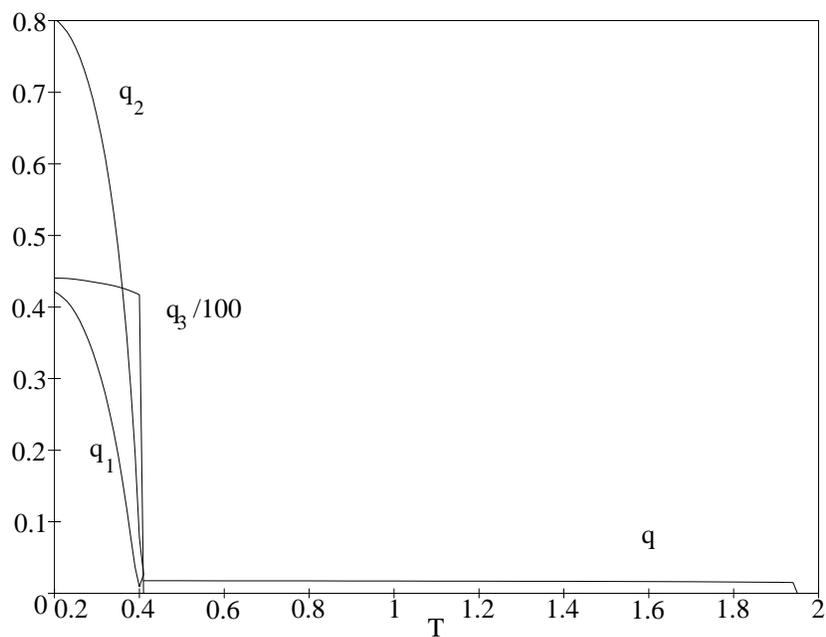}
\end{center}
\caption{
\label{fig4}
\interlinia
Temperature dependence of the twist parameters $q$ and $q_{i}$.
In the uniaxial cholesteric phase (between $T=1.94$ and $T=0.41$)
we have $q>0$.
In the biaxial cholesteric phase (below $T=0.41$) all $q_{i}$ 
are positive and $q_{1}<q_{2}<q_{3}$.}
\end{figure}

For simplicity in our calculations we took into account only the 
dependence on the order parameters $\langle F_{00}^{(2)}\rangle$
and $\langle F_{22}^{(2)}\rangle$.
We assumed the square-well potential energy of interactions with
$R_{SW}=2$.
Note that  $\sigma_{0}$ determines the length scale,
whereas $\epsilon$ determines the energy scale.
The elastic constants will be expressed in $\epsilon/\sigma_{0}$ and
$\epsilon/\sigma_{0}^{2}$, 
temperatures will be given in $\epsilon/k_{B}$.
We assumed that molecules are similar to ellipsoids 
with three different axes
$(\sigma_{0}/2)\times (\sigma_{0})\times (2\sigma_{0})$.
We used $V/N=8\sigma_{0}^{3}$ and our set of the molecular parameters is
\begin{eqnarray}
& \sigma_{1}=\sigma_{0}/2,
\sigma_{2}=\sigma_{0}/2,
\sigma_{3}=-\sigma_{0}/4, &
\nonumber \\
& \sigma_{4}=\sigma_{5}=0,
\sigma_{6}=\sigma_{7}=\sigma_{8}=\sigma_{0}/10. &
\end{eqnarray}
In our system, on decreasing the temperature we meet the first-order
transition to the uniaxial cholesteric phase at $T=1.94$
and the second-order transition to the biaxial cholesteric phase
at $T=0.41$. The temperature dependence of the order parameters
$\langle F_{00}^{(2)}\rangle$ and $\langle F_{22}^{(2)}\rangle$
is presented in Fig. \ref{fig1}. 
The order parameters were used to calculate the temperature dependence 
of the elastic constants. It appeared that the factors $\xi$ and $\eta$
introduce very strong variations with temperature and it was more 
comfortable to present the elastic constant divided
by those factors as shown in Figs. \ref{fig2} and \ref{fig3}. 
The corresponding twist parameters $q$ and $q_{i}$ 
are shown in Fig. \ref{fig4}.

In the uniaxial cholesteric phase there is one chiral constant ($K_{0}$)
and three nonchiral ones ($K_{1}$, $K_{2}$, and $K_{3}$).
The twist parameter $q$ describing the pure twist is decreasing
with temperature. For some potential energies one can obtain also the 
increasing temperature dependence, more often present in cholesterics.
We have also the accidental equality $K_{1}=K_{3}$ caused by the
omission of the order parameters $\langle F_{\mu\nu}^{(j)}\rangle$
with $j$ greater than 2. Typically, for rodlike molecules we get
the inequalities $K_{2} < K_{1} < K_{3}$.

In the biaxial cholesteric phase we have three chiral constants
($K_{11}$, $K_{22}$, and $K_{33}$) and 12 nonchiral ones.
On decreasing the temperature, $K_{0}$ splits into $K_{11} > K_{22}$ and
$K_{33}$ becomes nonzero. 
$K_{1}$ splits into $K_{1212} > K_{2121}$,
$K_{3}$ splits into $K_{3232} > K_{3131}$,
$K_{2}$ splits into $K_{2222} > K_{1111}$,
$K_{1122}$ becomes a new independent constant.
Other new independent constant are four positive 
$K_{1133}$, $K_{1313}$, $K_{2323}$, $K_{3333}$ 
and one negative $K_{2233}$.
They are about two orders of magnitude smaller then the previous ones.
The twist parameters $q_{i}$ are quickly increasing 
with the decreasing temperature and satisfy inequalities 
$q_{1} < q_{2} < q_{3}$.

\subsection{Binary phase of uniaxial molecules}

We mixed the prolate molecules interacting via the soft core potential
with $m=12$. The set of molecular parameters is
\begin{eqnarray}
& \sigma_{0}^{AA}=\sigma_{0},\ 
\sigma_{1}^{AA}=0.2\sigma_{0},\ 
\sigma_{2}^{AA}=-0.2\sigma_{0},\ 
\sigma_{6}^{AA}=-0.1\sigma_{0}, 
&\nonumber \\ &
\sigma_{0}^{BB}=\sigma_{0},\
\sigma_{1}^{BB}=0.3\sigma_{0},\ 
\sigma_{2}^{BB}=-0.2\sigma_{0},\ 
\sigma_{6}^{BB}=0,     &
\end{eqnarray}
where $\sigma_{0}$ denotes some molecular length and
$V/N=4.471\sigma_{0}^{3}$. 
The elastic constants will be expressed in 
$10^{-6}\epsilon/\sigma_{0}^{2}$ and 
$10^{-6}\epsilon/\sigma_{0}$,
the temperature in $10^{-6}\epsilon/k_{B}$.

For simplicity reasons we assume that only one phase is present. 
It corresponds to the stable solution with the minimum free energy.
The possibility of phase separation will not be investigated.
We checked the stability against perturbations of nematic symmetry 
and the stability against deformations of the phase.

\begin{figure}
\begin{center}
\includegraphics[scale=0.5,angle=270]{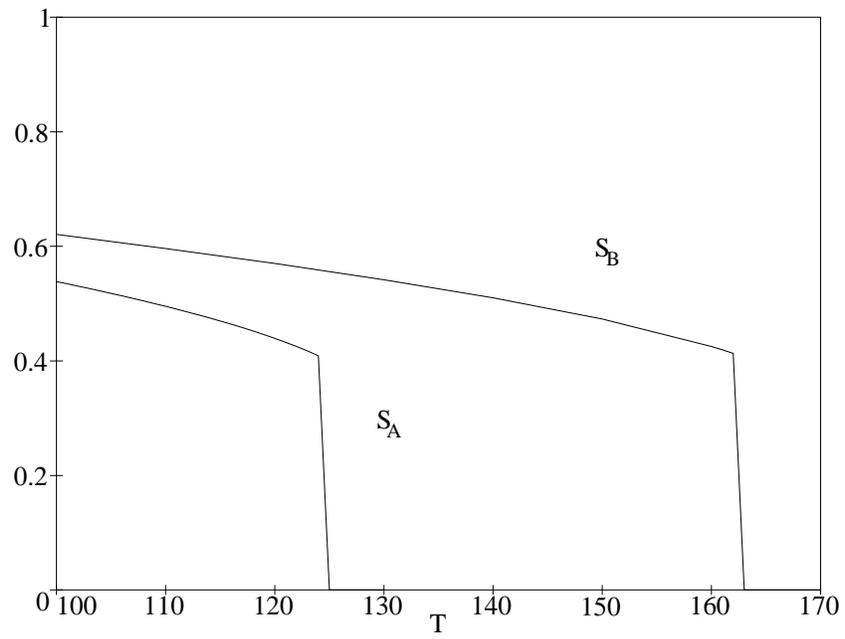}
\end{center}
\caption{
\label{fig5}
\interlinia
Temperature dependence of the order parameters 
$S_{A}=\langle P_{2}\rangle_{A}$ (Cho\-les\-teric) and
$S_{B}=\langle P_{2}\rangle_{B}$ (Nematic) for the unary systems.}
\end{figure}

\begin{figure}
\begin{center}
\includegraphics[scale=0.5,angle=270]{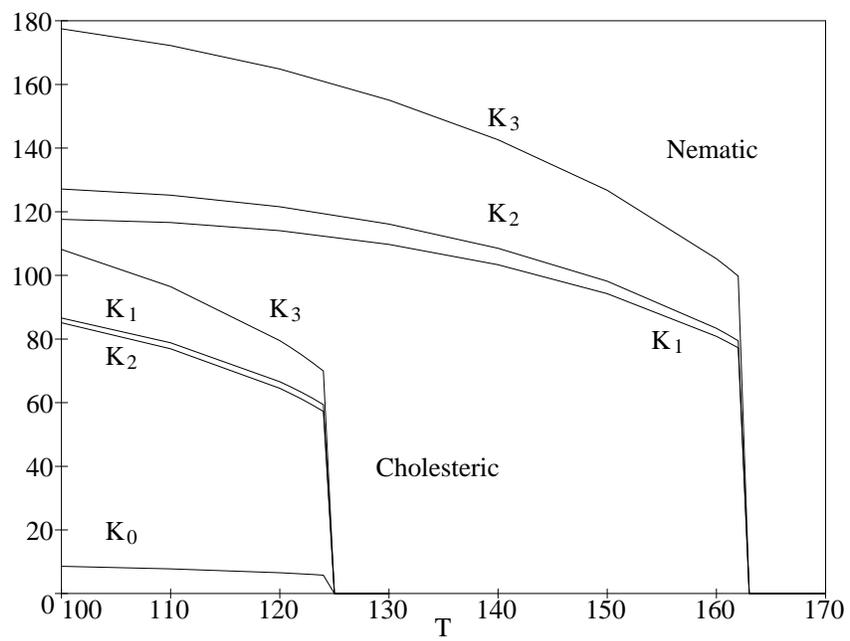}
\end{center}
\caption{
\label{fig6}
\interlinia
Temperature dependence of the elastic constants for the unary systems.}
\end{figure}

The molecules $A$ and $B$ in the unary phase form the uniaxial 
cholesteric and nematic phase, respectively. 
The transition temperatures of unary 
systems from the isotropic to the uniaxial cholesteric or nematic
phase are $T_{A}=124$ and $T_{B}=162$. 
The temperature dependence of the order parameters 
$\langle P_{2}\rangle_{I}$
and the elastic constants for unary systems are shown 
in Figs. \ref{fig5} and \ref{fig6}. 
The pitch of the system of 
molecules $A$ is a falling concave function of the temperature.

\begin{figure}
\begin{center}
\includegraphics[scale=0.5,angle=270]{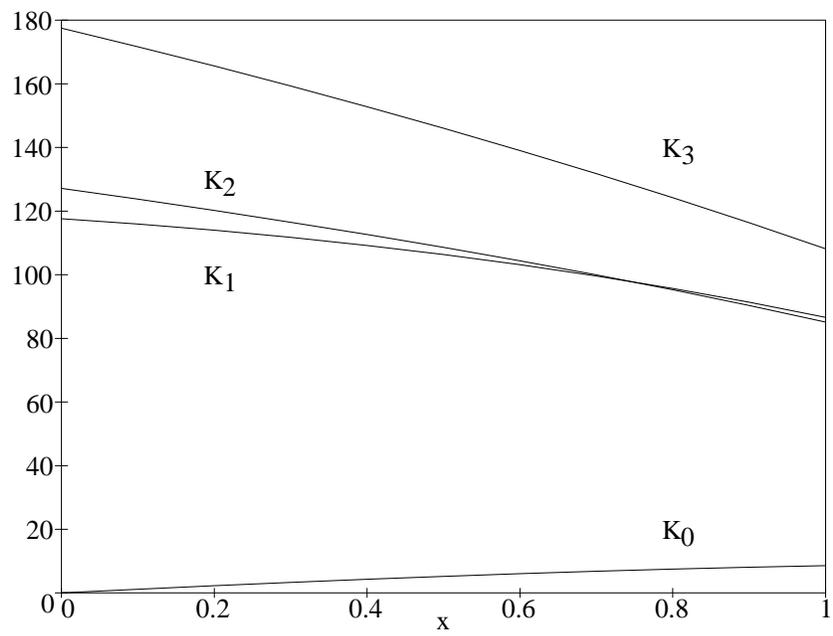}
\end{center}
\caption{
\label{fig7}
\interlinia
Elastic constants vs composition of the mixture at the temperature 
$T=100$ $(x=N_{A}/N)$.}
\end{figure}

In the considered range of temperatures the isotropic or uniaxial 
cholesteric phase was present in the mixture. 
The transition temperature $T_{C}$ of the binary system from the 
isotropic to the uniaxial cholesteric phase
can be described approximately as
\begin{equation}
T_{C}=\sum_{I=A,B}x_{I}T_{I}.
\end{equation}
We performed a detailed analysis of the mixture 
at the temperature $T=100$ where the uniaxial cholesteric phase 
was present for all concentrations.
The order parameters and the elastic constants are continous 
and concave functions of the concentration $x_{A}$. 
The dependence of the elastic constants on the concentration 
is plotted in Fig. \ref{fig7}.
The twist of the mixture is approximately described by the equation
\begin{equation}
\label{eq:q=sumxq}
q=\sum_{I=A,B}x_{I}q_{I},
\end{equation}
where $q_{I}$ are related to the unary systems at the same temperature.
The additivity rule (\ref{eq:q=sumxq}) is satisfied for most mixtures 
between cholesterol derivatives 
\cite{[1993_de_Gennes]} 
but there are some exceptions 
\cite{[1969_Adams_Haas_Wysocki]}. 
Note that in general $q_{I}$ may be positive or negative.

\section{Conclusions}
\label{sec:summary}

In this paper we developed the statistical theory for the uniaxial
and biaxial cholesteric phases.
Rigid molecules interacting via two-body short-range forces
were assumed.
We derived the microscopic formulas for the elastic constants 
and the pitch of the phase.
In order to calculate the values of the elastic
constants one needs the one-particle distribution function and
the potential energy of the molecular interactions. 
It is showed that the elastic constants determine the twist of the phase.
The obtained stability condition is to our knowledge not to be found
in the literature. In the case of the uniaxial cholesteric
or nematic phase it has the simple form $K_{1}<3K_{2}$. 
It seems that it is satisfied for all known substances.

Our theory was applied to the unary and binary systems of molecules 
similar to ellipsoids with three different axes. 
Thanks to the Corner potential energy the elastic constants
were expressed as a finite series of the order parameters.
Apart from this the role of the temperature is more transparent:
the function $B_{2}(T)$ determines the order parameters; 
the functions $B_{3}(T)$ and $B_{4}(T)$ are closely related 
to the chiral and nonchiral elastic constants, respectively.
The temperature dependence of the order parameters, 
the elastic constants and the twists 
in uniaxial and biaxial cholesteric phase was obtained.

The results concernig the nonchiral elastic constants $K_{ijkl}$
are the same as in the case of nematic phases described 
in \cite{[1997_Kapanowski]}.
We focus on the new results on the chiral elastic constants $K_{ii}$.
During the transition from a uniaxial to a biaxial cholesteric phase 
$K_{0}$ splits into $K_{11}$ and $K_{22}$, whereas $K_{33}$ becomes 
nonzero.
Below a certain temperature they satisfy inequalities
$K_{11}<K_{22}<K_{33}$. The corresponding twist parameters satisfy
similar relations $q_{1}<q_{2}<q_{3}$. 
Note that those relations depend on the type of molecular interactions 
and do not have to be always satisfied.

In the binary mixture it appeared that the transition temperatures 
and the twist of the mixture approximately satisfy 
the simple additivity rules.

Our theory is a starting point for further researches. 
It is desirable to go beyound a low density limit where the Mayer
function is replaced with a better approximation of the direct
correlation function $c_{2}$. The theory can be generalized
to the case of more component liquid crystalline mixtures.
We are also waiting for the experimental data on the biaxial nematic
and cholesteric phases and the elastic constants as, according
to Kini and Chandrasekhar, experiments are feasible
\cite{[1989_Kini_Chandrasekhar]}. The experiments can verify
the theoretical predictions and provide new problems.

\section*{Acknowledgment}

This research was supported by the State Committee for Scientific
Research (KBN), Grant No. 2P 03B05415.

\end{document}